\documentclass[aps,preprint,groupedaddress]{revtex4}
\usepackage{graphicx}
\usepackage{amssymb}
\usepackage{amsmath}
\usepackage{epstopdf}
\usepackage{longtable}

\begin{document}


\title{Red blood cells and other non-spherical capsules in shear flow: oscillatory dynamics and the tank-treading-to-tumbling transition}
\author{J.M. Skotheim$^*$ and T.W. Secomb$^+$}
\affiliation{$^*$Center for Studies in Physics and Biology, The Rockefeller University, New York, NY 10021 \\
$^+$Department of Physiology,
University of Arizona, 
Tucson, AZ 85724}

\date{\today}

\begin{abstract}

We consider the motion of red blood cells and other non-spherical microcapsules dilutely suspended in a simple shear flow.  Our analysis indicates that depending on the viscosity, membrane elasticity, geometry and shear rate, the particle exhibits either tumbling, tank-treading of the membrane about the viscous interior with periodic oscillations of the orientation angle, or intermittent behavior in which the two modes occur alternately.  For red blood cells, we compute the complete phase diagram and identify a novel tank-treading-to-tumbling transition at low shear rates.  Observations of such motions coupled with our theoretical framework may provide a sensitive means of assessing capsule properties.

\end{abstract}
%


\maketitle

Human red blood cells suspended at low concentrations in steady simple shear flow have been observed to exhibit two types of motion.  
While cells suspended in a low viscosity media tumble continuously, cells suspended in a fluid with sufficiently high viscosity 
exhibit `tank-treading' motion  \cite{F78}.  
Here, we use the term tank-treading to describe a cell that maintains almost constant shape and orientation in the laboratory frame, 
but whose membrane circulates around the interior much like the motion of a tank tread.
Keller and Skalak \cite{KS82} analyzed the motion of a tank-treading viscous ellipsoid and concluded 
that the behavior depends on the ratio of the viscosities of the inner and outer fluids and was independent of the shear rate.
However, it has recently been observed that red blood cells oscillate about a fixed angle while tank-treading and that tumbling can be induced by lowering the shear rate \cite{VFLA03}.
Furthermore, the dynamics of synthetic microcapsules in simple shear flow have been shown to depend not only on the viscosity ratio, but also on the shear rate: at high shear rates, the capsule surface tank-treads 
about the interior while the orientation oscillates; at low shear rates, the capsule tumbles  \cite{WRL00, WRL01}. 
These observations are unaccounted for by current theory and motivate the present work.
In this letter, we provide a unified theoretical framework for analyzing the motion of both non-spherical capsules and red blood cells in simple shear flow
and attribute the observed shear rate dependent behavior to a periodic variation in the elastic energy during tank-treading.

Here, we model the cell or capsule as an ellipsoid of viscosity $\mu'$ contained in an elastic membrane and immersed in a fluid of viscosity $\mu$.  
In the frame of the ellipsoid having axes of length $a_i$, the surface is at the position $\frac{x_i^2}{a_i^2}=1$. 
The external flow unperturbed by the ellipsoid is a simple shear flow of rate $\dot\gamma$ in the laboratory reference frame (see figure 1a). 
We assume that material elements in the membrane move along elliptical paths in planes parallel to the $x_1$-$x_2$ plane, with velocity field as in \cite{KS82}.
Then, the position of each membrane element is defined by a phase angle $\phi(t)$ and the tank-treading frequency is $\partial_t\phi(t)$. 
When $\phi = 2\pi$ a point on the surface has returned to its starting point.  
Our formulation extends the analysis of \cite{KS82} for a viscous ellipsoid to include the effect of an elastic membrane and we refer the reader to 
the literature for the derivations of several equations unrelated to elastic membrane.
Torque balance yields an equation for the rate of change in the orientation of the ellipsoid, $\theta$, in the laboratory frame \cite{KS82}
\begin{equation}
\label{dtheta}
\partial_t\theta = -\frac{\dot\gamma}{2} - \frac{2 a_1 a_2}{a_1^2+a_2^2}\partial_t\phi +\frac{\dot\gamma}{2} \frac{a_1^2-a_2^2}{a_1^2+a_2^2}\cos2\theta
\end{equation}
The tank-treading frequency, $\partial_t\phi$, is found by balancing the rate of work done by the external fluid on the
ellipsoid with the internal dissipation and the rate of change of the elastic energy stored in the membrane. 
The rate of work done by the external fluid on the ellipsoid, $W_p$, and the internal dissipation, $D$, are \cite{KS82}
\begin{eqnarray}
\label{workrate}
W_p = V \mu [f_2 (\partial_t\phi)^2 + f_3 \dot\gamma \partial_t\phi \cos 2\theta], \\
\label{dissipation}
D = V \mu' f_1 (\partial_t\phi)^2,
\end{eqnarray} 
where $V$ is the volume of the ellipsoid and $f_i$ are functions of the ellipsoid geometry defined in the endnote \cite{extraeq}.

In a series of recent experiments on red blood cells in a transient shear flow, Fischer \cite{F04} showed that after relaxation from deformation the rim of a red blood cell is always formed by the same part of the membrane.  
This shape memory could not be eliminated by continuous deformation for periods up to 4 hours, showing that the effect is not due to visco-elastic relaxation.
These results imply the existence of an elastic energy that has a minimum when the membrane is in static equilibrium. 
Since an element displaced from the rim can return 
to either side of the rim depending upon its proximity \cite{F04}, the elastic energy must pass through two minima during a tank-treading revolution.
Similarly, for the case of a polyamide capsule, the unperturbed shape is a slightly non-spherical ellipsoid and we assume that the elastic energy during tank-treading has the same 
$\pi$-rotational symmetry as the undeformed shape.
Consequently, we take elastic energy to have the form
\begin{equation}
\label{elenergy}
E = E_0 \sin^2(\phi).
\end{equation}
Fischer's experiments  \cite{F04} can also be used to estimate the size of the elastic energy change, $E_0$.  
When the shear flow is stopped, the rate of work done by the external fluid $W_p$ vanishes. 
The slender geometry of the red cell allows us to neglect dissipation in the external fluid in this analysis so that 
the rate of change in the elastic energy (\ref{elenergy}) balances the internal 
dissipation (\ref{dissipation}) so that 
$$
0 = E_0 \sin(2\phi) \partial_t\phi + V \mu' f_1 (\partial_t\phi)^2.
$$
The volume of a red blood cell is $V \approx 7\cdot10^{-17} \hbox{m}^3$ and $f_1 \approx 15/4$.
The internal viscosity $\mu'$ is due to a combination of the cytoplasm viscosity and the membrane viscosity.
The dissipation due to the membrane viscosity has been shown experimentally to be between 2 and 4 times the dissipation in the cytoplasm \cite{TSR84,F80}, and we choose the
value $\mu' = 4 \mu_{interior} = 4\cdot 10^{-2} \hbox{Pa} \cdot \hbox{s}$ corresponding a membrane dissipation equal to three times the dissipation in the cytoplasm.
Fischer's data show the maximum tank-treading frequency to be approximately $\partial_t\phi \approx  0.25 \hbox{s}^{-1}$ so that 
\begin{equation}
\label{elasticenergy}
E_0 = V \mu' f_1 |\partial_t \phi|_{max}  \approx 10^{-17} \hbox{J}.
\end{equation}
We can now estimate the amount of strain corresponding to the stored energy $E_0$.
The properties of the red blood cell membrane are determined by both the phospholipid bilayer and the adjacent protein network.  
The high resistance to change in membrane surface area insures that shear (area preserving) deformations dominate.
The resistance to shear and the shape memory are most likely due to the underlying protein network \cite{F04}.
For a membrane of shear modulus $E_s \approx 6\cdot10^{-6}\hbox{N/m}$ \cite{HW87} covering a surface $S \approx 1 \cdot10^{-10}\hbox{m}^2$ and undergoing a shear strain $u$ the elastic energy is approximately 
$E_0 \approx \frac{E_s u^2 S}{2}$. 
Equation (\ref{elasticenergy}) then yields the characteristic shear strain $u \approx  0.2$, a reasonable value.

Equating the rate of work by the external fluid (\ref{workrate}) with the internal dissipation (\ref{dissipation}) and the rate of elastic energy storage (\ref{elenergy}) yields 
\begin{equation}
V \mu (f_2 \partial_t\phi^2 + f_3 \dot\gamma \partial_t\phi \cos 2\theta) = V \mu' f_1 (\partial_t\phi)^2+E_0 \sin(2\phi) \partial_t\phi,
\end{equation}
which can be solved for the tank-treading frequency 
\begin{equation}
\partial_t\phi = \frac{\dot\gamma\,f_3}{f_2-\lambda f_1} (U_e \sin 2\phi - \cos 2\theta),
\end{equation}
where $\lambda = \mu'/\mu$ is the viscosity ratio and
$U_e = \frac{E_0}{V \mu \dot\gamma f_3}$ is the ratio of the change in the elastic energy to the work done by the external fluid during a rotation, i.e., how stiff the cell or capsule is 
relative to the forcing from the external shear flow.  
Scaling time, $t = T/\dot\gamma$, yields the dimensionless system of equations for $\theta$ and $\phi$
\begin{eqnarray}
\partial_T\phi = \frac{f_3}{f_2-\lambda f_1} (U_e \sin 2\phi - \cos 2\theta), \nonumber \\
\label{system}
\partial_T\theta = -\frac{1}{2} - \frac{2 a_1 a_2}{a_1^2+a_2^2}\partial_T\phi + \frac{1}{2} \frac{a_1^2-a_2^2}{a_1^2+a_2^2}\cos2\theta.
\end{eqnarray}
The system's behavior is now completely determined by $U_e$, $\lambda$, and the geometric ratios $a_1/a_2$, $a_1/a_3$.
Although observations of microcapsules have shown that the capsule geometry oscillates periodically during tank-treading \cite{WRL01,CO93}, 
periodic changes in axes length have little impact on solutions of (\ref{system}).

A simple way to understand the relationship between the tank-treading and oscillation periods is to 
 expand (\ref{system}) using $U_e$ as a small parameter: $\phi = \phi_0 + U_e \phi_1$, $\theta = \theta_0 + U_e \theta_1$.  
The $O(U_e^0)$ solution recovers the case studied by Keller and Skalak \cite{KS82}, where the tumbling transition is induced via an increase in the viscosity ratio and the steady solution 
is described by the constant orientation angle $\theta_0$ and the tank-treading frequency $\partial_T \phi_0 = \omega$.  
The $O(U_e^1)$ perturbation to this steady state is $\theta_1 = c \sin (2 \omega T - \delta)$, where the constant $c$ and the phase lag $\delta$ between the elastic energy and the orientation 
angle depend on the specific parameters of the system.  
Hence, $\theta = \theta_0 + U_e c \sin (2 \omega T - \delta)$ and
the two minima of the elastic energy in each tank-treading cycle cause the system to oscillate twice about a fixed angle in the time it takes a surface element to tank-tread back to its initial position.  
This effect is illustrated in figures 1b and 1c and solutions of (\ref{system}) show that it is also true for large amplitude oscillations.
Slightly non-spherical polyamide microcapsules \cite{WRL00,WRL01} and numerical simulations of slightly non-spherical liquid capsules enclosed by elastic membranes \cite{RP98} in shear flow have shown oscillations at frequencies twice the tank-treading frequencies.

The transition to tumbling behavior at low shear rates occurs when the fluid shear stress acting on the particle is no longer sufficient to force the membrane to tank-tread up the elastic energy gradient.
The surface velocity decreases and the capsule 'solidifies' and begins to tumble.
In contrast to the tank-trading-to-tumbling transition for a purely viscous drop described by Keller and Skalak \cite{KS82}, in which there is no dependence on the shear rate, 
the transition described here depends on the shear rate since $U_e \sim 1/\dot\gamma$.  The transition point is a function of the system parameters: $U_{ec}(\lambda,\frac{a_1}{a_2},\frac{a_1}{a_3})$.
Figure 2 illustrates the tank-treading-to-tumbling transition for the geometric ($a_1/a_2 = 4$, $a_1/a_3 = 1$) and elastic ($E_0 =10^{-17} \hbox{J}$) parameters of a red blood cell immersed in a viscous fluid ($\lambda=1.5$) being forced at a variable shear rate. 
 
Beyond the critical shear rate the system exhibits type I intermittent behavior \cite{PM80,ML80}, i.e. a series of oscillations 
followed by a tumble followed by another series of oscillations and so on.  
This transition is most easily understood by studying the return map onto the line $\phi = 0$ in the $\theta$-$\phi$ plane modulo $\pi$.  For the initial condition 
$\theta = \theta_0$, $\phi =0$ the trajectory crosses the $\phi=0$ line again at $F(\theta_0)$, 
thus defining the return map.  Periodic trajectories appear as fixed points in the return map
$F(\theta_0) = \theta_0$ and the tank-treading-to-tumbling transition occurs via a saddle-node 
bifurcation in which an unstable limit cycle and a stable limit cycle coalesce and disappear.  
Near the bifurcation $U_e-U_{ec} \ll 1$ and 
we expand the return map about $\theta^*$, where $F(\theta_0) - \theta_0$ is smallest, so that
$$
F(\theta_0) = \theta_0 + \eta (U_e-U_{ec})  + \xi (\theta_0-\theta^*)^2,
$$
where $\xi$ and $\eta$ are constants.  Taking $\theta_n$ to be the angle after $n$ applications of the map $F$, we see that in the
vicinity of the constriction, where $F(\theta_0) \approx \theta_0$ and  $\xi (\theta_0-\theta^*)^2 \ll 1$,
the system exhibits nearly periodic oscillations and $\theta_n \sim n \eta (U_e-U_{ec})$.  
When  $\xi (\theta_n-\theta^*)^2 \sim \eta (U_e-U_{ec})$ the system has left the constriction and intermittent behavior occurs.  
Hence, the number of iterations of the map prior to the intermittent behavior scales as 
$$
n \sim 1/\sqrt{ (U_e-U_{ec})},
$$
as seen in figure 4.  As $U_e$ is increased further the system enters a regime where the intermittent behavior consists of a series of tumbles followed by a tank-treading revolution.  
Finally for $U_e > U_{ec*}$ the system fails to exhibit full tank-treading rotations and simply tumbles.  
In this case, the relevant return map is onto the line $\theta = 0$ in the $\theta-\phi$ plane, so that for the initial condition 
$\phi = \phi_0$, $\theta =0$ the trajectory crosses the $\theta=0$ line again at $F(\phi_0)$.
This transition is also a type I intermittent transition caused by the coalescence of two limit cycles.  The phase diagram illustrating the different regimes is shown for both the
capsule and red blood cell geometries in figure 3.  The insets in figure 3 show the return map in different regions and two examples of intermittent trajectories.  

By modifying (\ref{elenergy}) to include visco-elastic membrane properties, our framework can also account for the 
shear flow induced oscillations observed for oil drops coated with a visco-elastic protein layer \cite{CO93,EFW05}.  
The details of the behavior discussed in this letter, particularly in and around the intermittent regime, are sensitive to parameter values.  
Therefore, mechanical characteristics of cells and capsules could potentially be deduced from observations of such behavior.

\acknowledgments

{\small We acknowledge support via the NSF DMR grant 9732083 (JS) and NIH grant HL034555 (TS).  We thank Manouk Abkarian for introducing us to this phenomenon and for helpful discussions.}

\begin{figure}
\includegraphics[width=13cm]{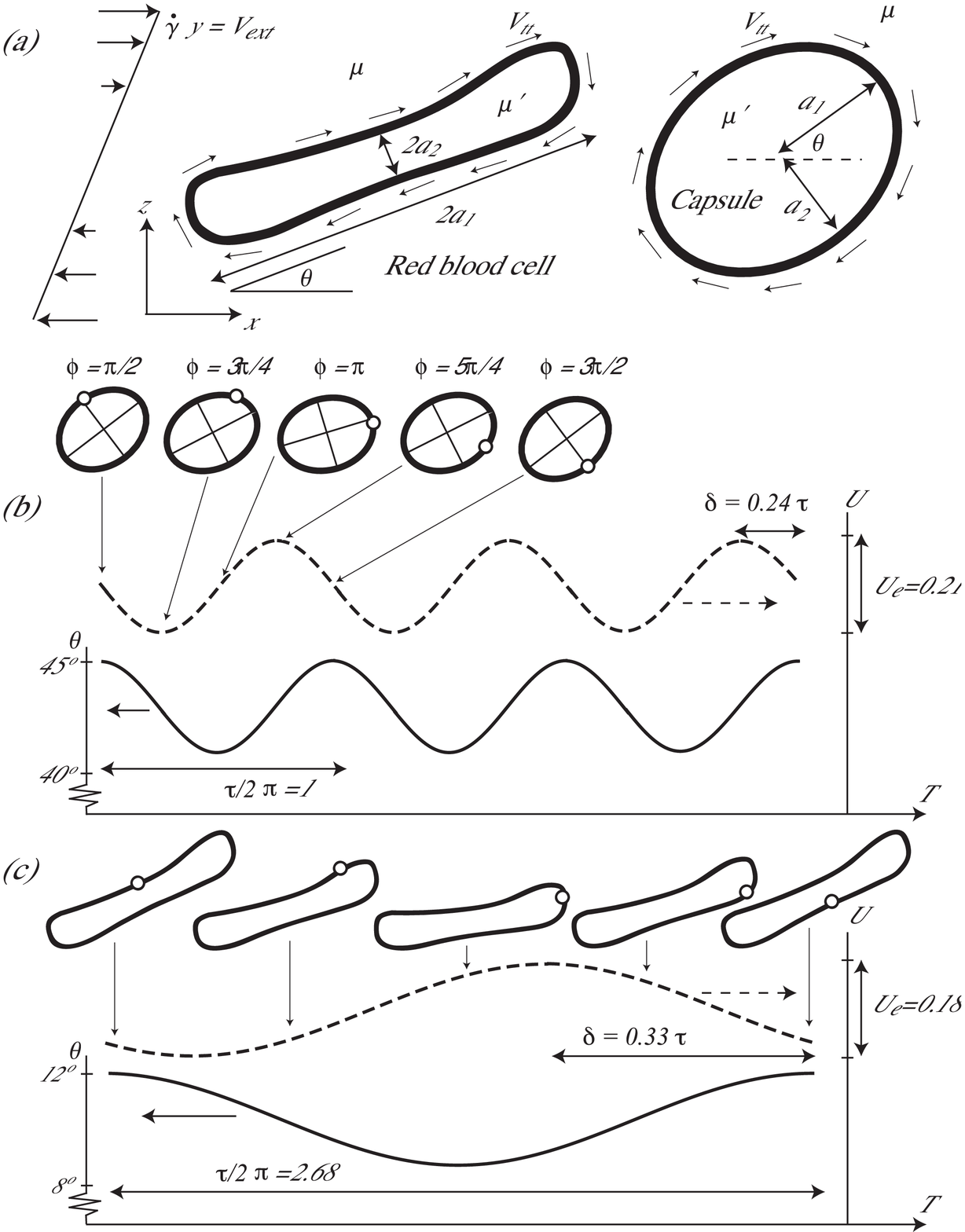}
\caption{\label{f.1} (a) Schematic diagram of a tank-treading red blood cell (left) and capsule (right) immersed in a fluid of viscosity $\mu$ that is being sheared at a rate $\dot\gamma$.  
The viscosity of the fluid inside the cell or capsule is $\mu'$ so that the viscosity ratio is $\lambda = \mu'/\mu$.  $\theta$ defines the orientation to the shear flow,
$V_{tt}$ is the surface velocity due to tank-treading, and $\phi$ defines the membrane orientation to the axes.  The red blood cell and capsule are modeled as ellipsoids of axes $a_i$.
$U_e$ is the dimensionless ratio of the resistance of the elastic membrane to deformation to the surface stress applied by the external fluid.
$\delta$ is the phase lag between $\theta$ and the elastic energy.  
(b) Capsule parameters: $a_1/a_2 = a_1/a_3 = 1.1$; $\lambda=0$; $\gamma=5$; $U_e=0.21$.  
(c) Red blood cell parameters: $a_1/a_2 = 4$, $a_1/a_3 = 1$;  $\lambda=1.5$; $\gamma=5$; $U_e = 0.18$.}
\end{figure}

\begin{figure}
\includegraphics[width=13cm]{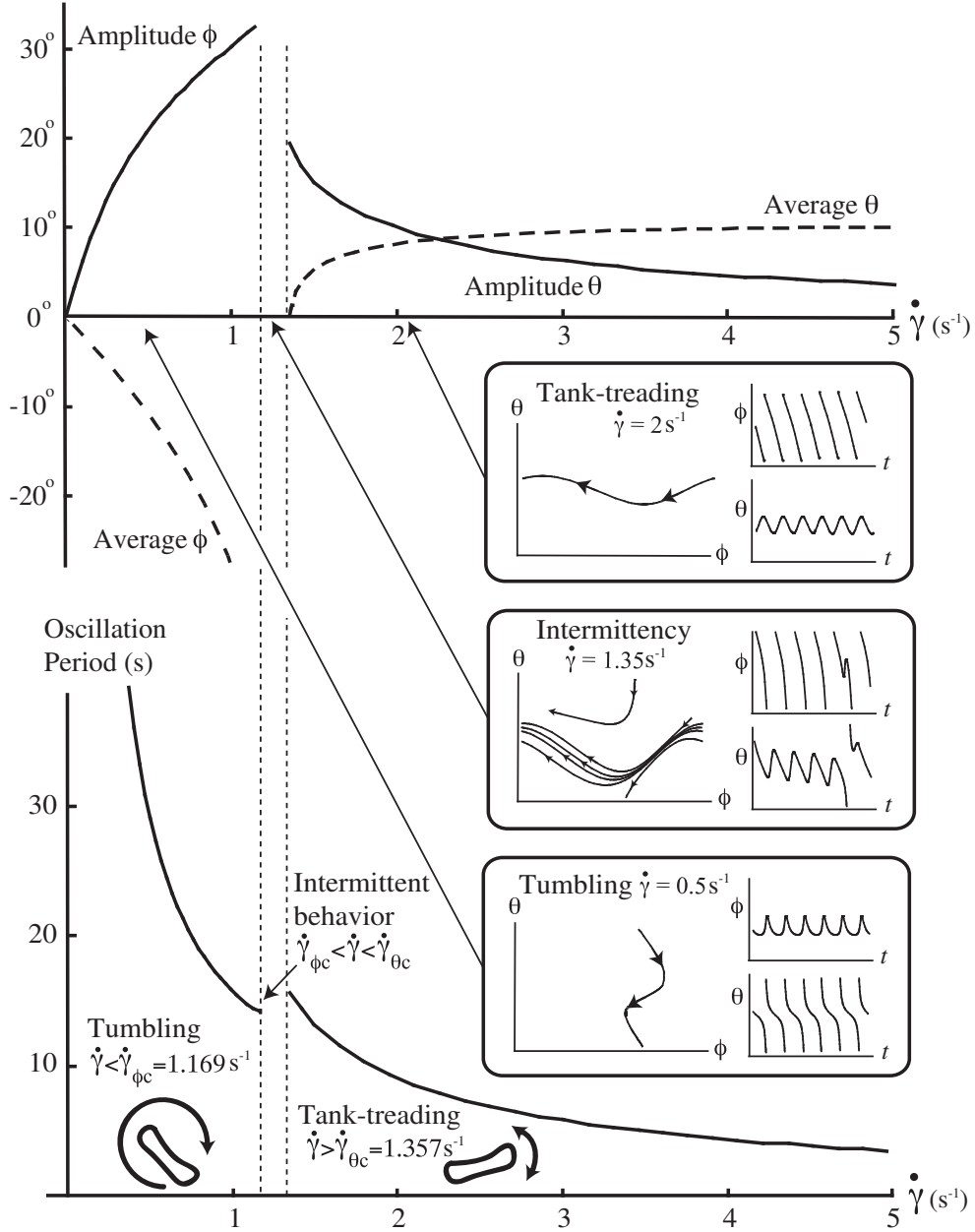}
\caption{\label{f.2} Red blood cell dynamics.  
For a red blood cell geometry ($a_1/a_2 = 4$, $a_1/a_3 = 1$) and elasticity \cite{F04} (see text) with a viscosity ratio $\lambda=1.5$ a novel transition from tank-treading to tumbling is predicted as the shear rate is decreased.
For $\dot\gamma > \dot\gamma_{\theta c}$ the membrane tank treads and the cell orientation $\theta$ oscillates with amplitude and average that depend on $\dot\gamma$.  
For $\dot\gamma < \dot\gamma_{\phi c}$ the cell tumbles and the orientation $\phi$ of the membrane relative to its unperturbed position oscillates.   
For $\dot\gamma_{\phi c} < \dot\gamma < \dot\gamma_{\theta c}$ the cell exhibits intermittent behavior, with a series of tank-treading revolutions followed by a tumble or a series of tumbles followed by a revolution.
Insets show the dynamical regimes for $\dot\gamma = 2, 1.35, 0.5$.}
\end{figure}

\begin{figure}
\includegraphics[width=12cm]{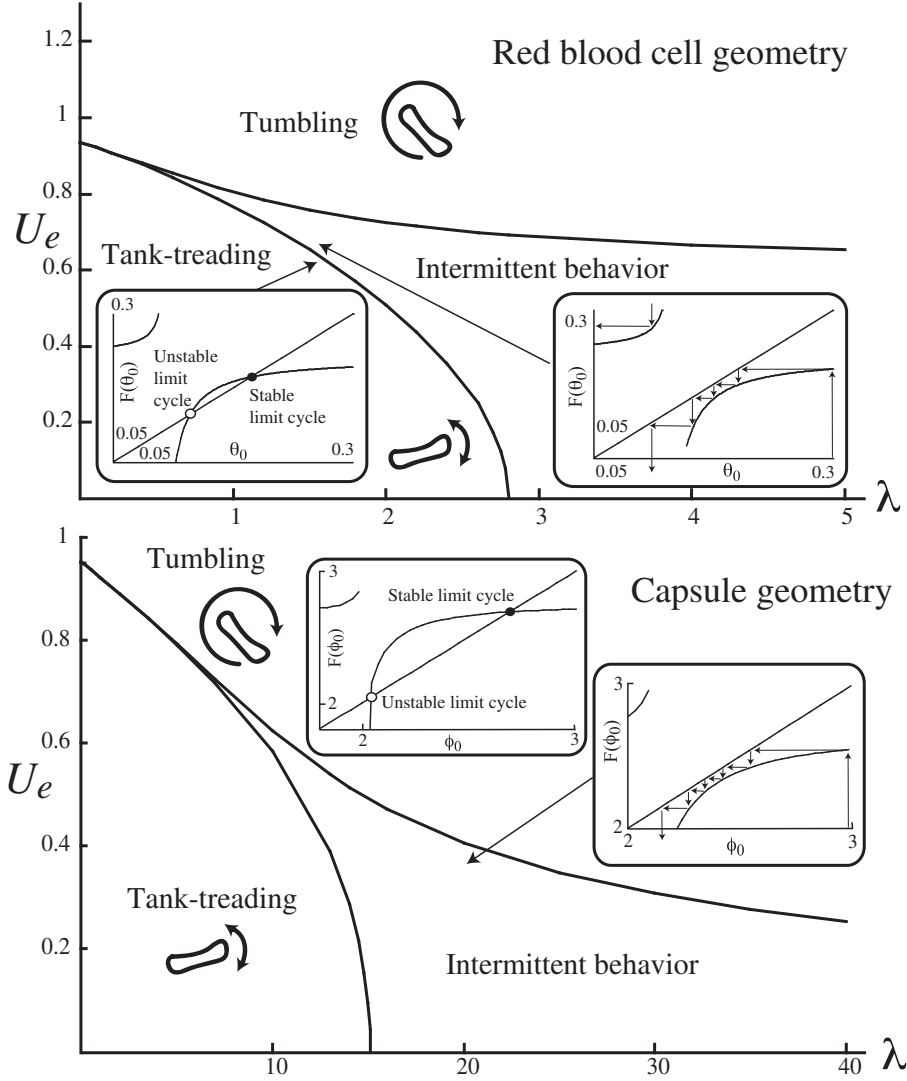}
\caption{\label{f.3} 
Phase diagram for the red blood cell ($a_1/a_2 = 4$, $a_1/a_3 = 1$) and capsule ($a_1/a_2 = a_1/a_3 = 1.1$) geometries. 
For a given geometry, the system behavior is determined by $U_e$, the dimensionless ratio of the elastic resistance to tank-treading divided by surface stress from the external flow,
and $\lambda$, the ratio of internal to external viscosity.
For the red blood cell geometry, the insets ($\lambda=1.5,~U_e=0.6557,~0.65$)
show how the return map
($F(\theta_0)$ gives the value of $\theta$ when the trajectory modulo $\pi$ beginning at $\theta=\theta_0,~\phi=0$ again crosses $\phi=0$)
changes as the system crosses the boundary between tank-treading and intermittent regions.  The stable and unstable fixed points of the return map coalesce in a saddle-node bifurcation and intermittent behavior ensues.
Similarly, for the capsule geometry a return map is shown for $\phi$ for $\lambda=20,~U_e = 0.6,~0.37$.  Both transitions occur for the two geometries investigated.}
\end{figure}

\begin{figure}
\includegraphics[width=7cm]{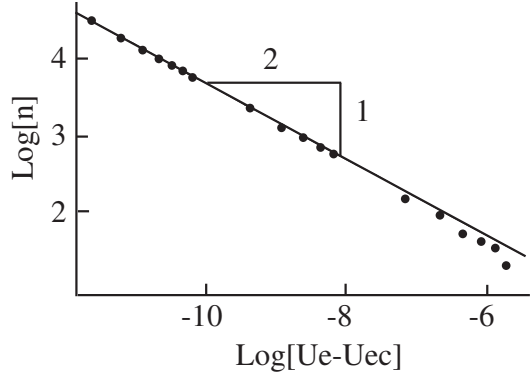}
\caption{\label{f.4} Type I intermittency: $n$ is the number of tank-treading oscillations per tumble;  
$U_e-U_{ec}$ is the difference between $U_e$ and its value at the boundary between the tank-treading and the intermittent regions.  
The plot is shown for $\lambda = 1.5$ and a red blood cell geometry ($a_1/a_2 = 4$, $a_1/a_3 = 1$).
The -1/2 slope on a log-log plot is indicative of Type I intermittency \cite{PM80} (see text), indicating the coalescence of two limit cycles, which appear as fixed points on the return map of figure 3.} 
\end{figure}

\end{document}